# Software search is not a science, even among scientists: A survey of how scientists and engineers find software

M. Hucka[a,*], M. J. Graham[b]

[a]*Department of Computing and Mathematical Sciences, Mail Code 306-15.*
[b]*Department of Astronomy, Mail Code 158-79. California Institute of Technology, Pasadena, California 91125, USA*

## Abstract

Improved software discovery is a prerequisite for greater software reuse: after all, if someone cannot find software for a particular task, they cannot reuse it. Understanding people's approaches and preferences when they look for software could help improve facilities for software discovery. We surveyed people working in several scientific and engineering fields to better understand their approaches and selection criteria. We found that even among highly-trained people, the rudimentary approaches of relying on general Web searches, the opinions of colleagues, and the literature were still the most commonly used. However, those who were involved in software development differed from nondevelopers in their use of social help sites, software project repositories, software catalogs, and organization-specific mailing lists or forums. For example, software developers in our sample were more likely to search in community sites such as Stack Overflow even when seeking ready-to-run software rather than source code, and likewise, asking colleagues was significantly more important when looking for ready-to-run software. Our survey also provides insight into the criteria that matter most to people when they are searching for ready-to-run software. Finally, our survey also identifies some factors that can prevent people from finding software.

*Corresponding author.
Email addresses: `mhucka@caltech.edu` (M. Hucka), `mjg@caltech.edu` (M. J. Graham)

## 1. Introduction

Software is critical to research (Hannay et al., 2009; Hettrick, 2014; Howison and Bullard, 2015; Howison et al., 2015; Ince et al., 2012; Katz et al., 2016; Katz and Ramnath, 2015; Morin et al., 2012; Stewart et al., 2013; Wilson, 2006), yet finding software suitable for a given purpose remains surprisingly difficult (Bourne, 2015; Cannata et al., 2005; Howison and Bullard, 2015; White et al., 2014). Few resources exist to help users discover available options or understand the differences between them (White et al., 2014). A recent study (Bauer et al., 2014) of developers at the Internet search company Google underscored the depth of the problem: the authors found the factor "most disruptive to the [software] reuse process" was "difficulties in finding artifacts." In other words, *even developers at Google have difficulty finding software.*

Searching the Internet with a general-purpose search engine has previously been reported to be one of the most popular approaches (Samadi et al., 2004; Umarji et al., 2008). Despite its popularity, this approach suffers from demonstrable problems. It requires devising appropriate search terms, which can be challenging for someone not already familiar with a given topic or who is not a native English speaker. Web searches also can yield dozens of viable candidates with little direct information about each, requiring the user to follow links and examine individual candidates—a time-consuming and tedious task. Finally, some questions cannot be answered through Web searches without substantial additional effort, such as what are the differences between candidate software tools. Other approaches to finding software, such as looking in the literature or asking on social media, suffer from still other problems such as the potential for incomplete or biased answers. The difficulty of finding software and the lack of better resources brings the potential for duplication of work, reduced scientific reproducibility, and poor return on investment by funding agencies (Cannata et al., 2005; Council, 2003; Crook et al., 2013; Niemeyer et al., 2016; Poisot, 2015; White et al., 2014).

One of the first steps to providing more effective resources for finding software is to understand factors that influence how users locate and select software today. However, most prior work on this topic has focused on *software developers* searching for *source code*; few studies included nondevelopers or asked how people look for ready-to-run software rather than source code. In addition, prior work has examined the use of *search systems* to find software, but not other options such as the use of catalogs. A variety of software catalogs exist today (e.g., Allen et al., 2012; Black Duck Software, Inc., 2016; National Aeronautics and Space Administration, 2016; Noy et al., 2009) and automated catalog generation has been explored (e.g., Kawaguchi et al., 2004; Linares-Vásquez et al., 2014; Linstead et al., 2009; Ossher et al., 2009; Tian et al., 2009; Ugurel et al., 2002; Yang and Tu, 2012), but there appears to be no direct study of users' preferences and use of software catalogs.

In an effort to understand these and other aspects of how people—specifically those in scientific and engineering disciplines—find software, in late 2015 we conducted a survey involving members of numerous mailing lists primarily in the fields of astronomy and systems biology. In this article, we report on five of the research questions addressed by our survey:

**RQ1** How do scientists and engineers look for ready-to-run software?

**RQ2** What criteria do scientists and engineers use when choosing ready-to-run software?

**RQ3** What information would scientists and engineers like to find in a catalog of software?

**RQ4** How do software developers in science and engineering look for source code?

**RQ5** What can prevent software developers in science and engineering from finding suitable source code?

This survey contributes to the body of research on discovery, search and reuse of software by people working in scientific and engineering disciplines.

Here, "reuse" is meant broadly and encompasses application reuse, component library reuse or source code reuse (Frakes and Fox, 1995; Gerard et al., 2007; Sim et al., 1998; Stolee et al., 2014). The survey results provide insights into people's current practices and experiences when searching for software in two different situations: looking for ready-to-run application software and looking for software source code. The outcomes also reveal the current role of catalogs in these situations, as well as people's preferences for information to include in catalogs. Overall, the survey results can inform the future development of improved resources to aid software discovery.

The remainder of this article is divided as follows. In Section 2, we overview related work. In Section 3, we describe our survey design and research methods, while in Section 4, we report our results. We discussion the results, implications and limitations in Section 5, and conclude with Section 6.

## 2. Related work

Some of our research questions have been examined or touched upon in other studies in the literature, though not all of the research questions have been previously addressed or examined in the same context or using comparable methods. We discuss related work in this section. We chose to organize the subsections below primarily by type of study; in Table 1, we provide an alternative view of the related work organized by our research questions RQ1–RQ5.

Table 1: Summary of other studies discussed in Section 2 in relation to our research questions. Italicized citations indicate works that are only partly comparable.

| # | Question | Related work |
|---|---|---|
| RQ1 | How do scientists and engineers look for ready-to-run software? | *Allen (1977)*<br>Bauer et al. (2014)<br>*Brancheau and Wetherbe (1990)*<br>Constant et al. (1996)<br>Eveland et al. (1994)<br>*Grossman et al. (2009)*<br>Joppa et al. (2013)<br>Lawrence et al. (2014, 2015)<br>*Matejka et al. (2009)*<br>Murphy-Hill and Murphy (2011)<br>Murphy-Hill et al. (2015)<br>*Rafique et al. (2012)*<br>Singer et al. (2014)<br>Twidale (2005)<br>Xiao et al. (2014)<br>Zmud (1983) |
| RQ2 | What criteria do scientists and engineers use when choosing ready-to-run software? | Joppa et al. (2013)<br>Lawrence et al. (2014, 2015) |
| RQ3 | What information would scientists and engineers like to find in a catalog of software? | *Bauer et al. (2014)*<br>*Gallardo-Valencia and Sim (2011a)*<br>*Lawrence et al. (2015)*<br>*Marshall et al. (2006)*<br>*Sadowski et al. (2015)* |
| RQ4 | How do software developers in science and engineering look for source code? | Bauer et al. (2014)<br>Berlin and Jeffries (1992)<br>Lawrence et al. (2014, 2015)<br>Marshall et al. (2006)<br>Orrego and Mundy (2007)<br>Sadowski et al. (2015)<br>Samadi et al. (2004)<br>Sim et al. (2012, 2011)<br>Singer et al. (2014)<br>Twidale (2005)<br>Umarji et al. (2008)<br>Umarji and Sim (2013)<br>Zmud (1983) |
| RQ5 | What can prevent developers from finding source code? | Frakes and Fox (1995)<br>Sim et al. (2012, 2011) |

## 2.1. Surveys examining how people find ready-to-run software (primary relevance: RQ1, RQ2)

Many surveys have examined software developers and search characteristics in the context of software *code* reuse, but few have examined how users—whether they are developers or not—go about locating and selecting *ready-to-run* application software (the topic of our RQ1 and RQ2). Our research uncovered only three reports of surveys that were not focused specifically on a software development context.

Joppa et al. (2013) surveyed 596 scientists working on modeling biological species distribution, and asked them what software they used and why they chose it. The reasons given by the respondents provide some insight into how the scientists found the software they used. In order of popularity, the answers that mentioned something about "how" were: "I tried lots of software and this is the best" (18% of respondents), "Recommendation from close colleagues" (18%), "Personal recommendation" (9%), "Other" (9%), "Recommendation through a training course" (7%), "Because of a good presentation and/or paper I saw" (4%), and "A reviewer suggested I use it" (1%). Surprisingly, none of the responses in Joppa et al. survey explicitly mentioned searching the Internet, although it is possible that some of the answers such as "I tried lots of software and this is the best" and "Other" subsumed the use of Web searches.

Lawrence et al. (2014, 2015) conducted a large survey about the use of science gateways by members of scientific communities. Several of their questions and results are pertinent to the topics of our own survey:

- They asked participants to indicate domains of expertise. The top five were "Physical and Mathematical Sciences" (30%), "Life Sciences" (22%), "Computer and Information Sciences" (16%), "Engineering" (16%), and "Environmental Sciences" (14%), though 16% did not indicate a domain. As we will discuss in Section 5.1, this is similar to the results of our survey.

- Lawrence et al. asked how people learn about and choose science gateways—a question related to our RQ1. They found that 78% indicated they learned about technologies from colleagues, 61% indicated conferences and other meetings as a source, 51% said publications, 38% said Web searches and speciality sites, 33% from students, and less than 10% from mailing lists or other methods such as magazine advertisements.

- Related to our RQ4, they asked software developers "How do you keep up to date with web-based technologies?", limiting answers to two choices from a predefined list and a free-text "Other" field. The three most popular answers were: using online communities via email or Web-based forums (47%), one's own development team (43%), and focused workshops (18%).

- In another question, Lawrence et al. (2015) asked participants "Assuming cost is not a factor, what are the most important factors you consider when adopting a new technology? Please select the three (3) most important factors in your decision-making process". Since this question had direct relevance to RQ2 in our survey, we include the full response results here:

  - "Documentation available" (49%)
  - "Ability to Adapt/Customize" (35%)
  - "Demonstrated Production-Quality Reliability" (31%)
  - "Availability of Technical Support" (30%)
  - "Open Source" (27%)
  - "Existing User Community" (20%)
  - "Interoperability with Other Systems" (20%)
  - "Availability of Support for Bug Fixes & Requests" (19%)
  - "Testimonials/User Ratings" (16%)
  - "Project Longevity" (13%)
  - "Licensing Requirements" (12%)
  - "Availability of Long-Term Maintenance" (11%)
  - "Reputation of Those Who Built the Software" (11%)

Finally, Murphy-Hill and Murphy (2011) and Murphy-Hill et al. (2015) studied social modes of software discovery, and focused on *unexpected* learning events in which a person does not realize they need a tool before learning about it. This is different from the focus of our study, which in Murphy-Hill et al.'s terminology is *purposeful discovery* or the deliberate seeking out of a solution when one is needed, a distinction that must be kept in mind when comparing these works because the latter is likely to involve different strategies and thus produce different responses. Nevertheless, their study examines how users and developers discover software (which is at the heart of our RQ1) and thus is appropriate to summarize here. They interviewed 18 programmers in industry and also performed a diary study with 76 software users, to explore how they learn about software from peers via social interactions. In the survey, they asked participants to recount situations in which they learned about a new software tool, and then the authors tallied the number of instances. The results were as follows:

- "Tool encounter" (where the person discovers the tool by exploring the interface of a development environment): 10/18 (60%)

- "Tutorial" (where the person is reading or watching a tutorial that mentions the software): 8/18 (44%)

- "Peer observation" (where the person observes someone else use a tool during a programming task): 7/18 (39%)

- "Discussion thread" (where the person finds out about a tool after reading about it in comments, forum discussions or email discussions): 4/18 (22%)

- "Written description" (where the person notices a tool mentioned in a website or publication): 3/18 (17%)

- "Twitter or RSS feed" (where the person finds out about a tool from someone or a site they read): 3/8 (17%)

- "Peer recommendation" (where someone observes the person during programming and suggests a tool): 1/18 (6%)

In the diary studies, Murphy-Hill et al. (2015) asked participants to rank the modes they considered most effective to least effective. Their participants consisted of students and technology workers; the students were further divided into computer science students (CS) and students of other majors (non-CS), and the workers into software developers and nondevelopers. The top three modes for each subgroup were as follows (estimated from the histogram in their Figure 5):

- CS students: (1) tool encounter, (2) Twitter or RSS feed, (3) tutorial

- Non-CS students: (1) peer observation, (2) peer recommendation, (3) tool encounter

- Software developers: (1) tool encounter, (2) Twitter or RSS feed, (3) tie between tutorial and written description

- Nondevelopers: (1) tutorial, (2) tool encounter, (3) peer observation

We relate the findings of Lawrence et al. (2015) and Murphy-Hill et al. (2015) further to our survey in the sections discussing our results for RQ1 (Section 5.2), RQ2 (Section 5.3), and RQ4 (Section 5.5).

## 2.2. Surveys examining the use of catalogs of software (primary relevance: RQ3)

Search engines are the predominant way that people look for software today, but as discussed in the introduction, there are limitations and drawbacks to using search. This has motivated the development of software catalogs (also sometimes called indexes, directories or registries), in which different software products are listed according to some classification scheme and often include detailed, structured information about the software products. Catalogs typically allow users to browse (and sometimes search as well) by various criteria (Allen and Schmidt, 2015; Katz, 2015; Marshall et al., 2006; Mena et al., 2006; White et al., 2014), and offer a more focused alternative to searching in general-purpose search engines. Numerous public software catalogs exist; most are domain/community-specific (e.g., Allen et al., 2012; Bönisch et al., 2013;

Browne et al., 1995; Gleeson, 2016; Goldberger et al., 2000; Hempel et al., 2016; Hucka et al., 2016; National Aeronautics and Space Administration, 2016; Noy et al., 2009; Shen, 2015), though some general catalogs also exist (e.g., Black Duck Software, Inc., 2016; Johansson and Olausson, 2016; Mario, 2016; SlashDot Media, 2016).

Surveys that examined people's approaches to finding software sometimes included questions about the use of catalogs, but they invariably concerned *whether* or *how often* users employed catalogs, not what information they sought in the catalog. For example, the survey by Lawrence et al. (2015) touches on the topic, since gateways often provide some kind of listing of accessible software; however, the authors do not report on how users employed the lists that may have been available. A survey by Marshall et al. (2006), discussed further in the next section, considered the question of whether users employed catalogs, but did not report the characteristics of those catalogs such as the information they contained. Similarly, the study by Bauer et al. (2014), also discussed in the next section, touched on browsing source code repositories and documentation, but does not describe the characteristics of the browsing facilities.

The general question of what information is useful for developers looking for software has been examined in some studies of programmer activities. For example, Gallardo-Valencia and Sim (2011a) conducted laboratory experiments to study the kinds of information used by software developers searching for source code on the web to solve programming problems. A study by Sadowski et al. (2015), described in more detail in the next section, examined the queries that developers used when searching their organization's source code repository; they found 26.5% of queries (out of 3,870) contained the operator for file path names (e.g., to restrict search within a particular software project) and 5.4% contained the operator to limit search to specific programming languages. However, the frequency of use of other operators is not reported in the 2015 paper.

Studies such as those of Gallardo-Valencia and Sim (2011a) and Sadowski et al. (2015) focus on search behavior and not use of catalogs. While they are undeniably important for their own goals, the studies provide only indirect and fragmentary insights into the types of information that users may find useful to provide in catalogs. We could find no directly-related work to compare to our RQ3, and thus we believe our results for RQ3 (discussed in Section 5.4) represent novel findings.

## 2.3. Surveys examining how developers find source code (primary relevance: RQ4, RQ5)

Most studies of how users find software have done so in the context of software development and the reuse of software source code. The types of reuse in these situations range from black-box reuse of programming libraries or other software components (i.e., reusing code "as-is"), to reusing (or simply studying) source code fragments; in addition, in programming contexts, many studies examined the reuse of other kinds of artifacts such as documentation, specifications, architectural patterns, and more.

Samadi et al. (2004) reported preliminary findings from a survey conducted by the NASA Earth Science Software Reuse Working Group. Their survey involved government employees and contractors in the Earth science community. Some results from the study are pertinent to our research question RQ4. First, on the topic of how people found reusable software artifacts, the following approaches were noted: (1) word of mouth or personal experiences from past projects, (2) general Web search engines (e.g., Google), and (3) catalogs and repositories. The authors report "Generic search tools (such as Google) were rated as somewhat important, whereas specialist reuse catalogs or repositories were not cited as being particularly important". Second, for criteria used to decide which specific components to choose, the authors report that "most respondents chose saving time/money and ensuring reliability as their primary drivers for reuse". The following additional considerations were noted: (1) "ease of adaption/integration", (2) availability of source code", (3) "cost of creating/acquiring alternative", and (4) "recommendation from a colleague". The authors found that (a) availability of support, (b) standards compliance, and (c) testing/certification, were "not ranked as particularly important".

Samadi et al.'s study was reprised in 2005 with a wider audience that included members of academia (Marshall et al., 2006). The authors reported that the survey produced essentially similar results to their 2004 survey. Related to our RQ5, Marshall et al. noted that the primary reason given by people for not reusing software from outside of their group was "they did not know where to look for reusable artifacts and they did not know suitable artifacts existed at the time." For those who *did* engage in reuse, "personal knowledge from past projects and word-of-mouth or networking were the primary ways of locating and acquiring software development artifacts." On the topic of how people located software, Marshall et al. noted "the use of reuse catalogs and repositories was rated the most important method of increasing the level of reuse within the community."

In a different NASA study, Orrego and Mundy (Orrego and Mundy, 2007) examined software reuse in the context of flight control systems. They studied 63 projects using interviews, surveys and case studies. In interviews with 15 people, the difficulty of assessing the characteristics of software was stated as the most problematic aspect of reusing software, usually because of inadequate documentation for the software to be reused.

Umarji et al. (2008) and Umarji and Sim (2013) surveyed Java programmers in 2006–2007 to understand how and why they searched for source code. Using invitations to mailing lists and newsgroups, they solicited participation to fill out a Web survey, and received 69 responses. Several facets of the Umarji et al. study are especially pertinent to our own survey (notably RQ4):

- The found common starting points for searches to be (1) recommendations from friends, and (2) reviews, articles, blogs and social tagging sites.

- With respect to how developers conducted searches, the participants in the survey used the following, in order of popularity: (1) general-purpose search engines (87% of participants), (2) personal domain knowledge (54%), (3) project hosting sites such as SourceForge.net (SlashDot Media, 1999) (49%), (4) references from peers (43%), (4) mailing lists (23%), and (5) code-specific search engines such as Google Code Search (Google, Inc., 2006) (16%).

- With respect to the selection criteria used by developers to choose a solution, Umarji and Sim (2013) report that the most important factors were: (1) software functionality (78%), (2) type of software license (43%), (3) price (38%), (4) amount of user support available (30%), and (5) level of project activity (26%).

Sim et al. (2012, 2011) reported a four-component study that mixed surveys, laboratory studies, and field observations. For their initial exploratory survey, they used an online survey system to ask professional software developers about their habits when searching for source code on the web. A total of 69 participants responded; the majority programmed in Java, C++ and Perl, and most had experience working on small teams with 1–5 people. One of the survey questions has direct relevance to our RQ4: a closed-ended, multiple-choice question where participants were asked "Which information sources do you use to search for code?" The responses were as follows:

- "Google, Yahoo!, MSN Search, etc.": 90%

- Domain knowledge: 54%

- Sourceforge.net, freshmeat.net: 49%

- References from peers: 43%

- Mailing lists: 23%

- Code-specific search engines: 16%

Singer et al. (2014) examined how software developers active on GitHub use Twitter. They conducted an initial exploratory survey with 271 GitHub users (270 of whom said they develop software) and followed it up with a validation survey involving 1,413 GitHub users (1,412 of whom said they develop software). Their results have the following relevance to the topic of how people find and choose software (RQ1 and RQ4). Developers extend their knowledge of software (including new software tools and components) by asking and answering questions, participating in conversations, and following experts. This can lead to serendipitous discovery of reusable methods, software components and software tools. Singer et al. noted "Some developers mentioned that Twitter helps them find and learn about things that they would not have been able to search for themselves, such as emerging technologies that are too new to appear in web searches."

Bauer et al. (2014) studied software reuse practices by developers at Google. Several questions in Bauer et al.'s survey are pertinent to our RQ1, RQ4 and to some extent, RQ3:

- They asked subjects for their top three ways of sharing software components. They received 63 responses: common repository (97%), packaged libraries (34%), tutorials (31%), blogs (19%), email (9%), "I do not share artifacts" (3%), and "other" (3%).

- Bauer et al. asked about the preferred ways to find reusable software. They received 106 responses: code search (77%), communication with colleagues (64%), Web search (49%), browsing repositories (41%), browsing documentation (23%), "other" (8%), "code completion" (5%), code recommendation systems (3%), and tutorials (3%).

- They also asked "What do you do to properly understand and adequately select reusable artifacts?" and received 115 responses: interface documentation (72%), examples of usage on blogs and tutorials (64%), reviewing implementations (64%), reading guidelines (51%), exploring third-party products (28%), "other" (10%), and participating in training for third-party products (5%).

Sadowski et al. (2015) surveyed and analyzed search behaviors of 40 software developers at Google, Inc. They used a combination of surveys and search log analysis generated by 27 programmers, with the surveys deployed as a browser extension that directed developers to the survey when they accessed the organization's code search site. Their research questions focused on why and when programmers search, as well as the details of the search queries and sessions. As noted in the previous section, their study touches on our RQ3; their survey also touches on our RQ4, in that it shows developers use code search tools to find software in the context of developing software, although their study did not examine other options for finding software.

## 2.4. Other studies and other perspectives

A number of other surveys have examined code search by developers (e.g., Gallardo-Valencia and Sim, 2011b; Singer et al., 1997; Xia et al., 2017) or other questions surrounding code reuse by developers (e.g., Frakes and Fox, 1995; Morisio et al., 2002; Sojer and Henkel, 2010), but differences in the specific questions asked and differences in methodologies make it difficult for us to compare results directly. In some cases, questions in the present study included answer options that were examined more deeply as separate questions in other works. For instance, Frakes and Fox (1995) surveyed 113 people from 28 US organizations and one European organization; most respondents worked for companies involved in software and aerospace. The goal of their often-cited study was to ask questions about software reuse. Most of the questions concerned different topics than our RQ1–RQ5, but one of their questions is related to our RQ5. Specifically, our RQ5 asked people to indicate which problems may have led to participants being unable to reuse code and included the answer option "concerns about intellectual property issues", which is similar to the question "I'm inhibited by legal problems" in Frakes and Fox's survey. However, their answer options for this question used a Likert-like scale whereas our RQ5 only allowed for a yes/no response for each of several different answer options, which means we can only relate the results in general terms (Section 5.6).

In the context of studying the behaviors of software developers, other methods besides surveys have also been applied. The two main classes of non-survey techniques have been the analysis of search engine logs (Bajracharya and Lopes,

2009, 2012; Brandt et al., 2010, 2009; Ge et al., 2014; Jansen and Spink, 2006; Li et al., 2009; Teevan et al., 2004; Völske et al., 2015), and observational studies—often coupled with interviews—either in institutional settings or in laboratory environments (Banker et al., 1993; Brandt et al., 2009; Dabbish et al., 2012; Gallardo-Valencia and Sim, 2013; Huang et al., 2013; Pohthong and Budgen, 2001; Sherif and Vinze, 2003; Sim et al., 2013; Sim and Alspaugh, 2011; Sim et al., 2011). The differences in methods or the type of data obtained in these efforts again make it difficult for us to compare results directly. For example, the studies by (Sim et al., 2012, 2011) discussed in Section 2.3 also included field observations in which the authors observed and interviewed 25 developers at a small software company. Sim et al. asked the developers about the goals of their code searches they made, their expectations, how they evaluated the matches, and how they were going to use the source code. Participants' answers given in the Sim et al. 2012 paper show that the answers have relevance to our RQ5, but the paper does not provide every participants' answers or a tally of all types of reasons given, so unfortunately we cannot meaningfully evaluate whether our results for RQ5 agree or disagree with the reasons given by their participants.

Finally, some additional works have examined questions about finding or discovering software from within specific theoretical frameworks such as human-computer interaction, social learning theory, and diffusion of innovation.

*Software learnability* is an area of study in human-computer interactions that includes topics such as exploratory learning of software (Rieman, 1996), where users learn to use software through exploration and trial and error, and other topics in how various software elements can help users discover and understand how to use software and software features (e.g., Grossman et al., 2009; Matejka et al., 2009; Rafique et al., 2012). Much of the work in this area is focused on how features help people learn how to use software, and less about *finding* software in the first place; however, to the extent that the studies include an examination of social communications and the consequent opportunities for discovering new software, software learnability does share concerns with the present study. Understanding the relationships more precisely is an area for future work.

*Social learning theory* examines how humans learn by observing others (Bandura and Walters, 1963). An example of work related to software discovery from this perspective is over-the-shoulder learning (Twidale, 2005), which involves the study of situations where peers help each other use software applications and the learners purposefully ask for help from the teachers. Relatedly, Berlin and Jeffries (1992) studied apprentice relationships involving computer scientists, Eveland et al. (1994) studied people seeking help from computer systems help providers, and Constant et al. (1996) studied how people use electronic communications networks to get help. These and similar studies examine in greater detail certain specific mechanisms of discovering software that are subsumed but treated more superficially in our RQ1 and RQ4.

*Diffusion of innovation* theory seeks to explain how new ideas (including new technologies) spread (Rogers, 2010). This has been applied to software diffusion (e.g., Allen, 1977; Brancheau and Wetherbe, 1990; Xiao et al., 2014; Zmud, 1983). A substantial amount of this research focuses on the characteristics of people and social networks involved in dissemination of new technologies, for example characterizing the behaviors of early adopters versus late adopters or analyzing the properties of social connection networks. However, some works have examined different information channels used to spread ideas (e.g., Xiao et al., 2014; Zmud, 1983), which is similar to some of the answer options in our RQ1 and RQ4. One of the most comparable such works is the interview-based study by Xiao et al. (2014). In one of their questions, they asked software developers what security tools they had heard about and how they heard about them. Out of the 14 predefined options given to the participants, the options that were mentioned at least once by participants were as follows:

- "Coworker Recommendation": 18/42 (43%)

- "Required by Management": 6/42 (14%)

- "Conference": 6/42 (14%)

- "Technical Blogs and Websites": 4/42 (10%)

- "Online Forums and Discussion Boards": 4/42 (10%)

- "Security Team Recommendation": 4/42 (10%)

- "Security Tool Vendor": 4/42 (10%)

- "Security Tool's Official Website": 3/42 (7%)

- "Required by Customer": 2/42 (5%)

## 3. Survey design and methods

Our survey was designed to shed light on current practices and experiences in searching for software in two different situations: looking for ready-to-run software, and looking for software source code. Respondents did not have to be software developers themselves (although the results show that most were). We chose to use a Web-based survey because it is an approach that (1) is well-suited to gathering information quickly from a wide audience, (2) requires modest development effort, and (3) can produce data that can be analyzed qualitatively and quantitatively.

### 3.1. Instrument development

Following the practices of other surveys in computing (e.g., Kitchenham and Pfleeger, 2008; Varnell-Sarjeant et al., 2015), we designed the instrument iteratively and paid attention to the following points:

- Wording. We sought to make the questions clear and unambiguous, and avoid implying a particular perspective. We elaborated each question with explanatory text under the question itself.

- Relevance to user's experiences. We limited our questions to topics that could reasonably be assumed to be within the experiences of our audience.

- Contemporary circumstances. We tried to ground the questions by referring to real resources and specific software characteristics that we believe are applicable to computer users today.

- Ethics. We avoided questions that might be construed as being too personal or about proprietary policies at participants' place of work.

### 3.1.1. Initial generation of questions and answer options

We obtained several question ideas from other surveys that touch on similar topics. In particular, we borrowed several question ideas from the survey instrument used by Sim et al. (2012). For example, our question "How often do you search online for software source code?" is nearly identical to their first question save for the addition of one more answer option. Some of our answer options for RQ4 (Figure 10) were initially drawn from their question ten, then expanded based on the content of some other surveys such as Sim et al. (2011) (notably their Table II and III) and our own experiences. To generate the list of options we used for our RQ3 (Figure 9), we initially began with the list of options in Sim et al.'s question twelve, then removed some options (their answers "Cost/effort required to adapt or integrate" and "Time to close bugs") that were too specific to software development for the purposes of our RQ3 (which was meant to be answered not only by software developers), and finally, added other answer options based on ideas from Table 1 from the paper by Crowston et al. (2006) and our own experiences.

The surveys cited above concerned software developers, but we strove to generalize the questions for RQ1–RQ3 and the answer options to make them applicable to nondevelopers as well. For example, answer options such as using general web search engines, mailing lists, social media, and looking in the scientific literature are not resources that only software developers would use, and so we sought to write the questions in a suitable way.

### 3.1.2. Iteration with initial pilot survey

The first version of the survey instrument had twenty-five questions. We performed a small pilot survey with our immediate colleagues and their students (approximately twenty people contacted this way) as well as the three external advisors of the project that funded this survey. One of the external advisors is a faculty in computer science at University of California, Irvine; another is a data scientist at the NASA Jet Propulsion Laboratory; and the third is a researcher in computational biology at the University of Oxford (UK). The faculty member also requested her students to take the survey and send us detailed feedback.

Based on the feedback of approximately ten people, we adjusted the survey form. We removed four questions from the first version of the form because feedback indicated they were either too ambiguous or people complained they took to long to answer: (1) "How common is it in your work group to look for existing software rather than to create your own software for a given task?" (multiple choice), (2) "Suppose an organized software catalog were available, where you could list and search for existing software based on various criteria. In your opinion, what are the most important capabilities such a resource should provide?" (free text), (3) "Please describe one or two scenarios when you were looking for source code on the Internet" (free text) and (4) "In your work, approximately what fraction of your time is spent on the following activities?" (two-dimensional multiple choice grid). We also added a new question "In your work, on a typical day, approximately what fraction of your time involves using or interacting directly with software on a computer or other computing device?" In addition, we elaborated the text of questions that were deemed unclear by the people who gave us feedback. For example, instead of the final wording of the question in RQ3, we had initially used "Which of the following characteristics ideally should be described for every software listed in a software catalog or index?" Finally, we added additional answer options to several of the questions: (1) "In your current (or most recent) software development project, what is (or was) your primary responsibility?", (2) "what are some approaches you have looked for source code", (3) "Which programming and/or scripting language(s) have you had the most experience with", and (4) "If you searched and found source code in the past, what are some factors that may have prevented you from REUSING the source code you found?" [1]

---

[1]The survey system did not provide a way to put words in bold text, so we used capital-

### 3.1.3. Final survey instrument

The final survey form is presented as a supplementary file to this article. The instrument contained a total of twenty-two questions (of which five were administrative or personal questions, such as name and email address), and included conditional branches between sections so that the final number of questions actually seen by any given respondent depended on the answers selected to certain screening questions. There were five main groups of questions in the survey:

1. Basic demographic and general information, suitable for all respondents.

2. Questions for software users who have the freedom to choose software. This section was only shown if participants indicated that they have some choice in the software they use.

3. Questions for software developers. This section was only shown if respondents indicated that are engaged in software development.

4. Questions for software developers who search for source code. This was only shown if respondents indicated both that they are software developers and that they search for software source code.

5. Survey feedback. This section sought feedback about the survey itself.

Questions in section No. 2 of the survey aimed to establish the relative importance of different search criteria. Those in section Nos. 3 and 4 sought to characterize the experiences of the developer.

The survey form used a mixture of four types of questions: check boxes, pull-down selection menus, two-dimensional rating grids, and short-answer input fields. Some of the questions allowed answers on a nominal scale (for example, approaches used for finding software), some questions used an ordinal scale (for example, the importance of different considerations when looking for software), and some were open-ended questions asking for free-form text.

---

ization to emphasize that the question was about *reusing* software. The immediate previous question in the survey asked about *finding* software, and there we capitalized the word "find".

### 3.2. Administration

We used Google Forms (Google, Inc., 2015a) to implement the survey instrument. The version of Google Forms was the free edition made available as of September, 2015. We obtained prior approval for the survey protocol from the California Institute of Technology's Committee for the Protection of Human Subjects. The survey form itself included a Web link to a copy of the informed consent form for survey participation. The first question in the survey provided a clickable checkbox by which subjects had to indicate they had read the informed consent form and consented to our use of their responses to the survey. This was the only question in the survey that required a response; all other responses were optional.

We generated a URL (Uniform Resource Locator) for the survey form using Google Shortener (Google, Inc., 2015b), a service that produces shortened URLs and simultaneously provides an analytics facility tied to the URL. On September 1, 2015, we invited participation in the survey. As mentioned below, we advertised the survey on mailing lists and social media oriented to the astronomical and biological sciences, particularly to computational subcommunities within those domains. Recipients were free to participate if they chose. The introduction and instructions for the survey were brief. The survey had no express closing date.

### 3.3. Sampling plan

We used nonprobabilistic convenience sampling with self-selection. We advertised the survey on electronic mailing lists oriented to the astronomical and biological sciences: the *IVOA* mailing list (astronomy), a Facebook astronomy list, the mailing list `sysbio@caltech.edu` (systems biology), the forum `sbml-interoperability@googlegroups.com` (systems biology), the list `cds-all@caltech.edu` (departmental list), and our immediate work colleagues (totalling a dozen people). Taken together, the members are a mix of staff, students, and faculty working in academia, government laboratories, and industry.

Potential biasing factors in the results include those that are common to self-selected written surveys with convenience sampling: response bias (i.e., people who responded may have different characteristics than those who did not), coverage errors (i.e., the representation of participants may not be balanced across different subcommunities), and item bias (i.e., some questions may have been skipped intentionally or unintentionally). An additional possible source of bias is that the authors are relatively well-known within the subcommunities to which the survey was advertised, which may have influenced respondents.

### 3.4. Population sample

We analyzed the results obtained by December 31, 2015. We estimate the number of potential recipients of the mail announcements to be at least 2300. The number of completed survey forms was 69. As mentioned above, our survey URL was backed by an analytics facility; this provided the number of URL clicks, the referrer sources, and source geographical locations. According to this facility, the survey form was accessed 172 times. Using these three numbers, we can calculate the following:

1. Estimated access rate to survey form: approximately 7.5% (172/2300).

2. Estimated response rate: approximately 3% (69/2300).

Unfortunately, we cannot be certain of the actual number of recipients. While we can determine the number of addresses we contacted, some of the addresses on mailing lists may be obsolete or unreachable, or the recipients' electronic mail systems may have filtered out the mail messages. Thus, we can only estimate the response rate.

### 3.5. Analysis

Descriptive statistics were performed using custom programs written in the language Python (Perez et al., 2011; van Rossum and de Boer, 1991), version 3.4, in combination with the NumPy (Van Der Walt et al., 2011) package, version 1.10.4. The figures in this paper were generated with the help of the Python library Matplotlib (Hunter, 2007), version 1.5.1. The statistical analyses were performed with the help of the Python `scipy.stats` package (from SciPy version 1.0.0), specifically `ttest_ind` for the t-tests, and the Python `Statsmodels` package version 0.8.0 (Perktold et al., 2017), specifically `stats.multitest.multipletests` for the Benjamini-Hochberg method (Benjamini and Hochberg, 1995).

## 4. Results

### 4.1. Demographics

Our survey included several questions to gather general demographic information. One of the first questions was "What is your primary field of work?", with multiple choices and "Other" as the answer options. Figure 1 shows the answer choices and responses. Of 69 respondents, 57% identified themselves as working in the physical sciences, 46% in computing and maths, 28% in biological sciences and 7% in a range of others. Subjects could select more than one field, and participants made use of this feature: 17 respondents selected two fields of work, six selected three fields, and one indicated four fields of work.

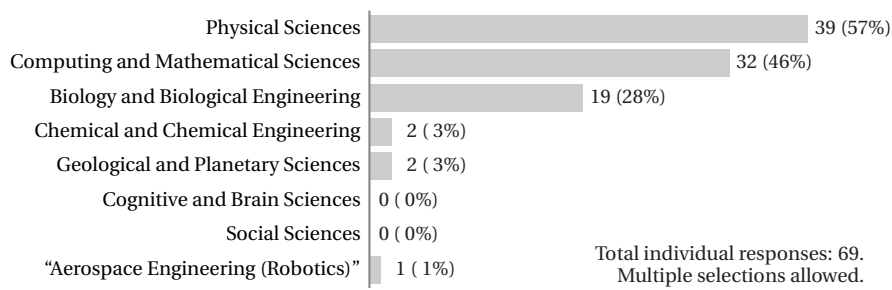

Figure 1: Respondents by discipline. This question offered the first eight predefined categories and a slot for write-in, free-text answers. Choices were nonexclusive. Some respondents wrote write-in answers that were clearly subsumed by one of the predefined categories; in those cases, we adjusted the totals appropriately. One response, "Aerospace Engineering (Robotics)", did not fit any predefined category; we included it as a true "Other" value.

To assess how computer-intensive our subjects' work activities are, the survey included the question "In your work, on a typical day, approximately what fraction of your time involves using or interacting directly with software on a computer or other computing device?" The answer options were in the form of a pull-down menu with values ranging from 0% (none) to 100% (all), in 5% increments. Note the question was not limited to time spent using technical software—respondents were free to interpret this broadly to mean any software used in a work context. Figure 2 provides a bar graph of the responses. The results show (Figure 2) that the overwhelming majority of our respondents spend over 50% of their day interacting with software. To quantify this further, assuming a typical 8 hour working day, we can conclude that 94% of participants regularly spent more than four hours of their day engaged with software, and 68% spent more than six hours.

As mentioned above, the overall motivation for the survey was to understand how scientists and engineers find software. An important consideration was whether subjects actually had a choice in the software they used. (The rationale for this is that if a person has no choice but to use software that is already provided or selected for them, then their answers to questions about how they find software would not be meaningful.) This motivated another question in the survey: "In your work, how much freedom do you usually have to choose the software you use?" Answers to this question were used to select subsequent survey questions: if a respondent answered "Never" to this question, then the remaining questions were skipped and people were shown the final survey page. Figure 3 provides the results for this question. The results show that every one

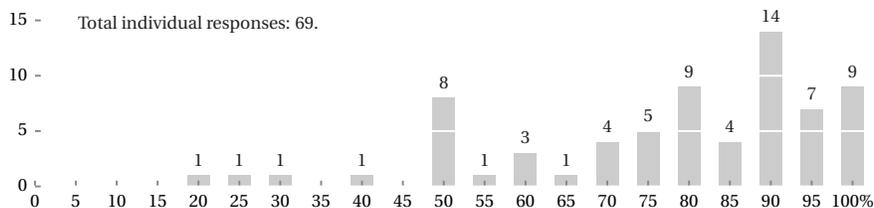

Figure 2: Bar graph of responses to the question "In your work, on a typical day, approximately what fraction of your time involves using or interacting directly with software on a computer or other computing device?"

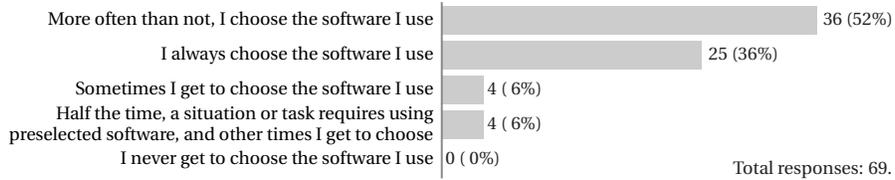

Figure 3: Responses to "In your work, how much freedom do you usually have to choose the software you use?".

of our respondents had some choice in the software they used.

In response to another question, "Are you involved in software development?", 56 (81%) answered "Yes" and 13 (19%) answered "No". The answer to this question controlled the display of additional questions of relevance to developers. Among the questions that were made available to the 56 who answered "Yes" were additional demographic questions. (Those who answered "No" were not shown the additional questions, and were instead taken to the final survey feedback page.) The first question for developers was "For how many years have you been developing software?" with a free-form text field for answers. We manually processed the 56 text responses to remove extraneous text and reduce them to numbers, and then tabulated the values. Figure 4 provides a histogram of the responses received for those who answered the question with an interpretable answer (55 out of 56).

Another question asked of those who indicated they were involved in soft-ware development was "In your current (or most recent) software development project, what is (or was) your primary responsibility?" It offered eight multiple choice items and a ninth "Other" choice with a free-form text field. The choices were nonexclusive: although we asked for people's primary responsibility, participants were free to choose more than one, and the explanatory text for the question indicated "If it is hard to identify a single one, you can indicate more than one below." Figure 5 provides a tally of the responses.

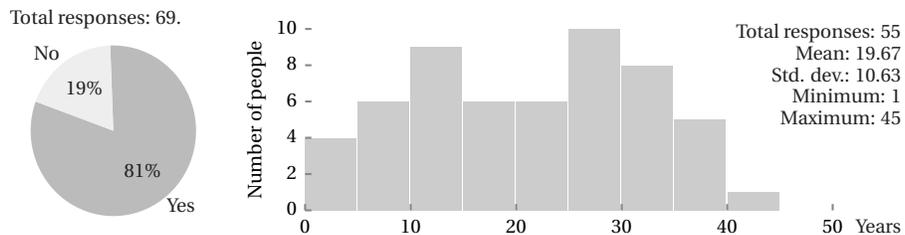

Figure 4: (Left) Responses to the question "Are you involved in software development?" (Right) Histogram plot of years that respondents have been developing software (for those who also answered "Yes" to the question of whether they were involved in software development).

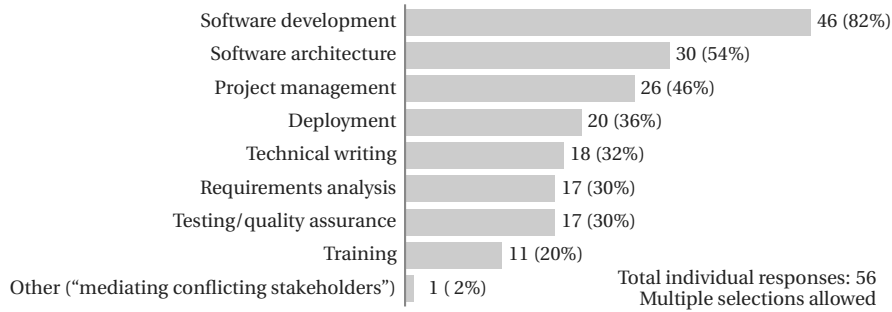

Figure 5: Responses to the question "In your current (or most recent) software development project, what is (or was) your primary responsibility?" This question was shown only to the 56 respondents who answered "Yes" to the question of whether they were involved in software development. This survey question offered the first eight predefined categories and an additional slot for free text under "Other"; only one respondent provide a value for "Other". Choices were nonexclusive.

The survey also posed the question, "What is the typical team size of projects you are involved with?" The form of the answers was again a set of multiple choice check boxes with an "Other" choice that offered a free-form text field. Answers were provided by all 56 respondents who answered "Yes" to the question of whether they were involved in software development (Figure 4), and none of the participants selected "Other". A total of 43 respondents (77%) selected "Small (1–5 people)", 12 respondents (21%) chose "Medium (6–25 people)", and 1 respondent (2%) selected "Large (more than 25 people)".

We also asked "Which programming and/or scripting language(s) have you had the most experience with?" in the set of questions only shown to those respondents who indicated they were involved in software development. We provided 22 predefined choices along with a free-text "Other" option. Choices were nonexclusive, and the elaboration under the question explicitly requested "Please select up to 3 languages which you have used the most." The top five responses were: Python (selected by 59% of participants), C (50%), Java (34%), shell scripting (32%), and C++ (27%).

Finally, our survey was designed to show a subset of questions of relevance to those developers who specifically indicated they searched for source code. The question "How often do you search online for software source code?" had six answer choices. If respondents chose any option *other* than "Never", they were shown further questions specific to searching for source code (pertinent to RQ4 and RQ5). The results are shown in Figure 6. Only one participant indicated "Never"; 55 respondents indicated they searched for source code at least some of the time.

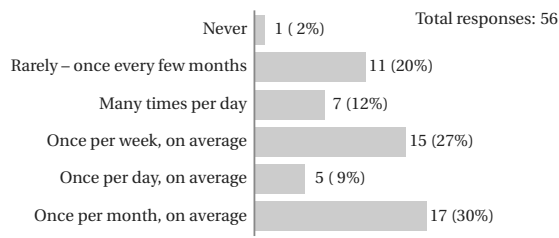

Figure 6: Responses to the question "How often do you search online for software source code?" Predefined answer choices were presented as the mutually-exclusive multiple choices shown on the vertical axis.

### 4.2. RQ1: How do scientists and engineers look for ready-to-run software?

As explained above, after the demographics questions, the remaining questions in the survey questions were only shown if respondents indicated they had a choice in selecting the software they used. As it turned out, all respondents indicated they have some choice in the software they use.

To assess how our respondents located or discovered ready-to-run software, we asked "When you need to find ready-to-run software for a particular task, how do you go about finding software?" The question provided multiple nonex-clusive answer choices together with a free-text "Other" option. Respondents were free to choose more than one answer. Figure 7 summarizes the results. The graph is sorted by the sum of responses across the separate subgroups. We computed p-values using an independent two-sample t-test assuming unequal sample sizes and unequal variances (Welch's t-test) for each option independently. Since the set of answer options constitutes a set of multiple comparisons, we also applied a correction for false positives suitable for situations involving multiple simultaneous tests. We chose the Benjamini-Hochberg method (Benjamini and Hochberg, 1995), and chose a false discovery rate of 10% as being appropriate to the goals of this exploratory study. In Figure 7, we indicated in bold font the p-values of those responses that are significant after correction.

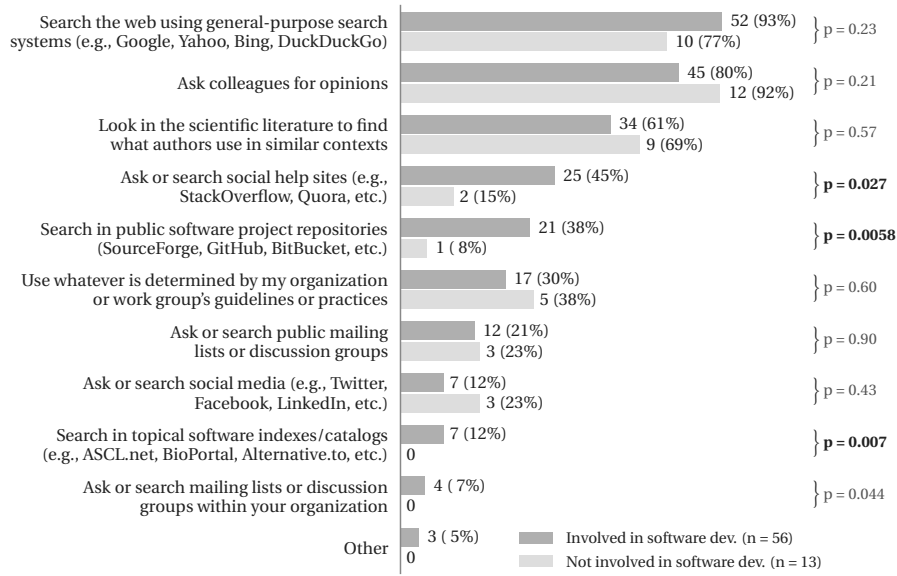

Figure 7: Responses to the question "When you need to find ready-to-run software for a particular task, how do you go about finding software?" Answer choices were nonexclusive. All 69 survey participants answered this question; results are subdivided according to respondents' answers to the question in Figure 4, where 56 people answered "Yes" to involvement in software development and 13 answered "No". Percentages are calculated by subgroup. The results of testing for unequal variances using Welch's t-test are given for differences between subgroups for each answer option. Bold typeface indicates differences that are significant after applying the Benjamini-Hochberg procedure (Benjamini and Hochberg, 1995) for controlling the false discovery rate to 10%.

### 4.3. RQ2: What criteria do scientists and engineers use when choosing ready-to-run software?

We sought to understand the selection and evaluation criteria that may come into play when our participants try to find ready-to-run software. We posed the question "In general, how important are the following characteristics when you are searching for ready-to-run software for a task?" For the answer options, we provided a two-dimensional grid with different predefined criteria as the rows, and values on a unipolar rating scale for the columns. The available values on the scale were "Rarely or never important", "Somewhat or occasionally important", "Average importance", "Usually of above-average importance", and "Essential". We chose this Likert-like scale by analogy to examples found in survey handbooks (e.g., Gideon, 2012, p. 101), with an adjustment to the wording of values to avoid implying finality. (E.g., instead of "Not important at all" we used "Rarely or never important", in recognition of the fact that users may find different criteria important on different occasions.) Then, in the analysis, to mitigate possible differences in interpretations of the values by different individuals, we ranked each criterion by using a percentage calculated as the sum of all three values of "Essential", "Usually of above-average importance" and "Average importance" divided by the number of possible responses (69). Figure 8 summarizes the results for the aggregate ratings by all participants.

To examine whether there is a difference in preferences between scientists and engineers who develop software and those who don't, we once again performed Welch's t-test between the responses of the two groups (developers and nondevelopers) for each possible characteristic. Though several p-values were below 0.05, none of the results were statistically significant after application of Benjamini-Hochberg correction for false positives (again using a false discovery rate of 10%). Thus, we have no sound basis to separate the responses of the two groups or rank their responses independently.

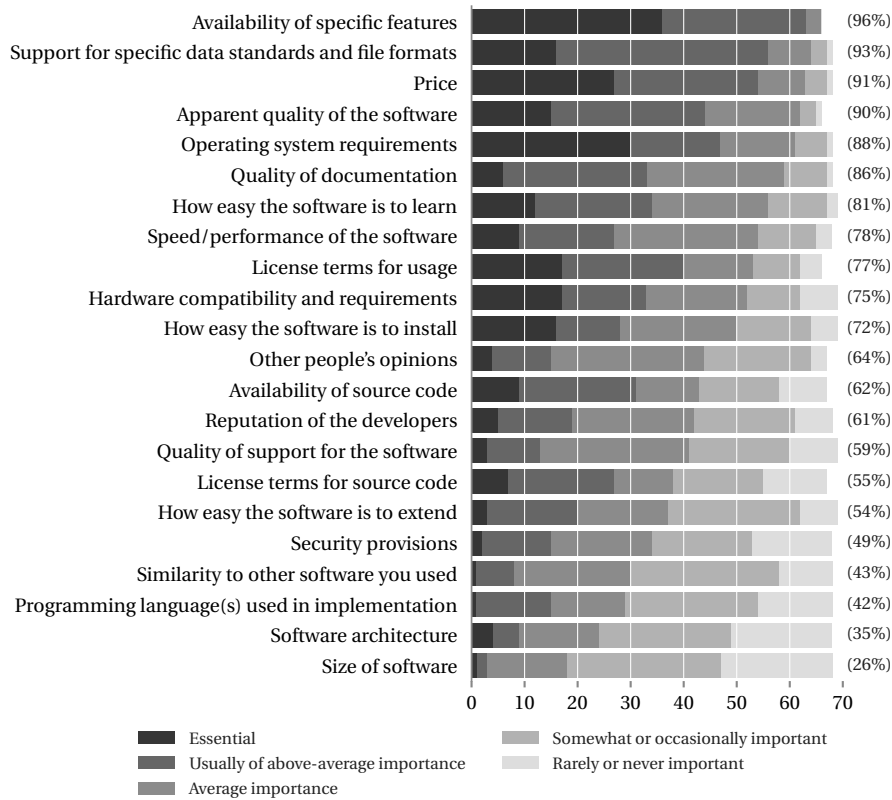

Figure 8: Responses to the question "In general, how important are the following characteristics when you are searching for ready-to-run software for a task?" All 69 respondents answered the question, but not all respondents chose to select an option for every possible characteristic. Responses are ranked by a percentage calculated from the sum of the number of "Essential", "Usually of above-average importance" and "Average importance" ratings for each option divided by the total number of possible responses (69).

## 4.4. RQ3: What information would scientists and engineers like to find in a catalog of software?

As discussed above (Section 2.2), a variety of software catalogs have been developed and many are available today. Being specialized and focused on software, they have the potential to be useful resources to augment or replace the use of general web search engines for finding software. However, the currently-available catalogs are highly heterogeneous in their features and the information they present to users (e.g., Allen et al., 2012; Bönisch et al., 2013; Browne et al., 1995; Gleeson, 2016; Hempel et al., 2016; Hucka et al., 2016; National Aeronautics and Space Administration, 2016; Noy et al., 2009; Shen, 2015), and as mentioned in Section 2.2, we found no studies of what information users wanted to see in such catalogs. This motivated our RQ3. To help inform the development of improved catalogs, we sought to determine what kind of information users find important to provide about software.

We posed the following question of all participants who indicated they had the freedom to choose software (not only those who indicated they developed software): "Suppose that it were possible to create a public, searchable catalog or index of software, one that would record information about software of all kinds found anywhere. What kind of information would you find most useful to include for each entry in such a catalog or index?" This question was in the section titled "Questions for software users" and followed two other questions about ready-to-run software. As with most other questions in our survey, we provided answer choices as nonexclusive multiple choices, with an additional free-text option titled "Other". All 69 participants to our survey replied to this question. Figure 9 summarizes the results. For greater insight, we separated the responses based on how individuals answered the yes/no question about being involved in software development (Figure 4). The graph in Figure 9 is sorted by sum of responses across developers and nondevelopers for each answer category.

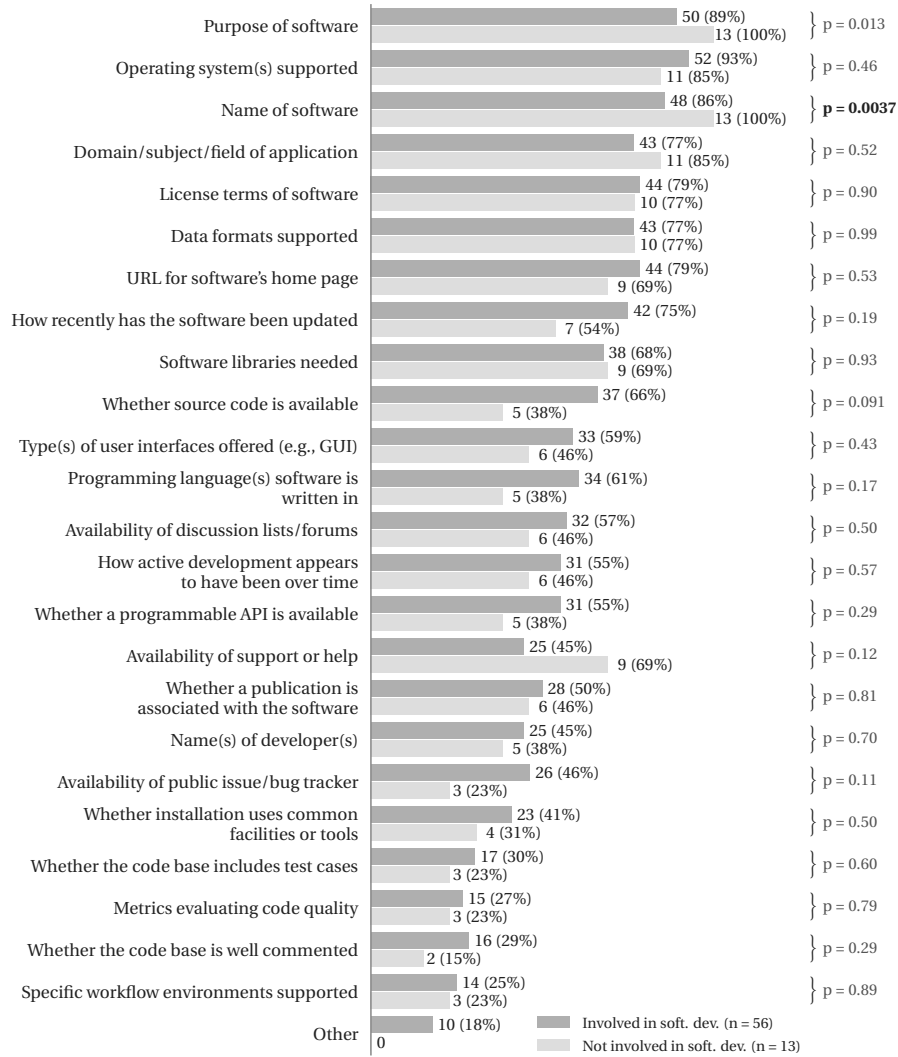

Figure 9: Answers to the question "Suppose that it were possible to create a public, searchable catalog or index of software, one that would record information about software of all kinds found anywhere. What kind of information would you find most useful to include for each entry in such a catalog or index?" There were 24 predefined items and a slot for free text under "Other". Choices were nonexclusive. All 69 survey respondents answered this question; results are shown subdivided according to participants' answers to the question in Figure 4 (left). The graph is sorted by totals; e.g., "Name of software" was the third most selected choice of developers and nondevelopers taken together. The results of testing for unequal variances using Welch's t-test are given for differences between subgroups for each answer option. Bold typeface indicates differences that are significant after applying the Benjamini-Hochberg procedure (Benjamini and Hochberg, 1995) for controlling the false discovery rate to 10%.

### 4.5. RQ4: How do software developers in science and engineering look for source code?

As explained in Section 4.1, the questions concerning searching for source code (RQ4, as well as RQ5 in the next section) were further gated by the question "How often do you search online for software source code?" A total of 55 participants indicated they searched for source code at least some of the time. These participants were shown additional questions, including "What are some approaches you have used to look for source code in the past?". Answer options were nonexclusive multiple choices, including an "Other" option with a field for free-text input. Figure 10 provides a summary of the results. This question was answered by all 55 participants who indicated that they searched for source code at least some of the time (Figure 6).

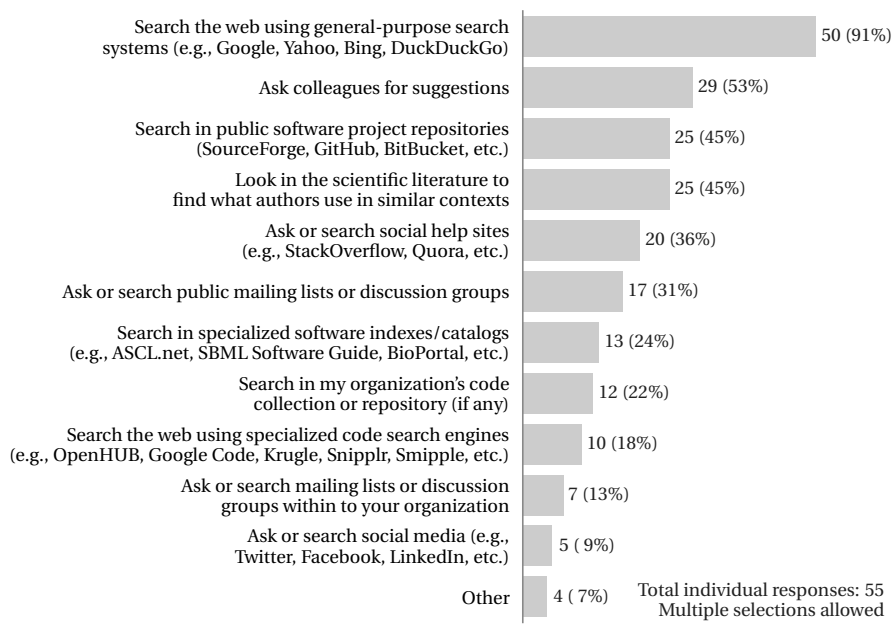

Figure 10: Responses to the question "What are some approaches you have used to look for source code in the past?" This question offered the first eleven predefined categories and an additional slot for free text under "Other". Answer choices were nonexclusive. This question was answered by the 55 respondents who self-identified as software developers.

### 4.6. RQ5: What can prevent software developers in science and engineering from finding suitable source code?

From our own experiences, we know a search for software can fail to find suitable candidates for a variety of reasons. This motivated our inclusion of another question in the survey: "What are some factors that have hindered your ability to FIND source code in the past?" [2] The question included a variety of nonexclusive predefined options, along with an "Other" option offering a free-text input field. The results are summarized in Figure 11.

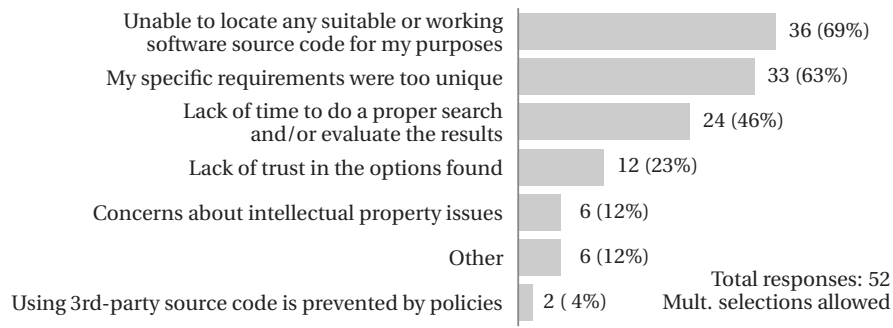

Figure 11: Responses to the question "What are some factors that have hindered your ability to FIND source code in the past?" This question offered the first six predefined categories and an additional slot for free text under "Other". Answer choices were nonexclusive. This question was answered by 52 of the 55 respondents who self-identified as software developers.

Six out of 52 respondents provided "Other" answers. Three of these were simply explanatory but did not add to the categories listed, two participants cited lack of documentation as a hindrance to either locating or evaluating software, and the third hindrance noted by a respondent was "Some scientific software is hidden from search engines as authors did not bother to put it online or make a small website for it."

---

[2] As mentioned previously, the survey system did not provide a way to put words in bold text, so we used capitalization to emphasize that the question was about *finding* software, to distinguish it from the next question in the survey which concerned *reusing* software.

## 5. Discussion

In what follows, we discuss the results in the same order and with the same headings as they were presented in the previous section.

### 5.1. Demographics

Most of our survey respondents worked in three areas: physical sciences, computing and mathematical sciences, and biology and biological engineering (Figure 1). The responses to the remaining demographic questions (Section 4.1) are consistent with prosaic expectations for the targeted scientific communities. We expected to reach computer-literate individuals, and due to the distribution channels used, we most likely reached those working in research environments. This is similar to other related studies such as that by Lawrence et al. (2015).

Languages such as Python and Java are popular in these settings, and our survey's numbers for languages are consistent with those of a recent Stack Overflow survey (Stack Exchange Inc., 2016) for "most popular technologies per dev type" for their participants who chose "Math & Data". Most of our respondents indicated involvement in software development, and their typical team sizes were small, with 77% being groups of 1 to 5 persons. This is common in scientific software development, especially in academia, and more generally in open-source communities (Sojer and Henkel, 2010). The fact that many respondents indicated they had multiple roles is also consistent—small teams generally require members to take on more than one role.

Among the 81% of the total 69 respondents who were shown the section for software developers, the median number of years of experience was 20 (see Figure 4). This suggests that the typical respondent is mid-career or part of the pre-mobile device generation. Of these, 70% (equal to 56% of the overall 69 respondents) indicated that they were also primarily responsible for project management and/or software architecture, which are traditionally more senior roles. The demographic data may thus indicate a bias in responses against more junior members of the respective communities, such as students and postdocs.

This is worth keeping in mind because junior members may have different search criteria and development experiences than more experienced colleagues. The possibility should be borne in mind when interpreting the survey results in the following sections. The cause of this distribution is unknown.

*5.2. RQ1: How do scientists and engineers look for ready-to-run software?*

Figure 7 summarizes the responses about how the highly-trained researchers in our study find ready-to-run software. For survey respondents involved in software development as well as those who are not, the top three most-selected choices were (a) using general search engines, (b) asking colleagues and (c) looking in the literature. There was no statistical difference in the popularity of these three approaches between the software developers and nondevelopers in our sample of scientists and engineers.

The differences that *were* significant were in the use of social help sites, software project repositories, and the use of software catalogs. Very few non-developers in our survey indicated they used social help sites such as Stack Overflow, and only one indicated they searched in software code repositories such as GitHub. While it is not surprising that software developers would be more familiar with these resources and thus recognize them as viable options for finding software, the results do indicate a difference in approaches used by developers versus nondevelopers in our sample of scientists and engineers.

There was also a statistical difference between the subgroups when it came to the use of domain-specific software catalogs; however, in absolute terms, even the scientific software developers did not seem to use them much. This last result is surprising. A possible explanation is that people may expect general search engines such as Google to index the domain-specific catalogs, and thus, that searching the former will subsume the latter. This *does* happen in practice: results from at least some of the domain-specific catalogs can easily be demonstrated to show up in Google search outputs, though using domain catalogs *directly* will usually produce fewer, more germane results. A second possibility is it reflects a belief that such resources are too narrowly focused in scope for their needs. A third possibility is that the results reflect ignorance of the existence of topical indexes. Future research should probe this issue further and seek to understand the reasons behind this result.

We were surprised by the overall low number of respondents in either subgroup indicating the use of general social media sites such as Twitter and Facebook. Other studies found a higher proportion of use (Bik and Goldstein, 2013; Singer et al., 2014), which led us to expect a similar outcome here, but the overall ranking of social media search in Figure 7 is quite low.

Finally, the write-in answers for "Other" revealed a category of options we did not anticipate: the use of network-based software package installation systems such as the systems available for the different Linux operating system distributions. In retrospect, this is an oversight in our list of predefined categories—the package management systems do offer some search capabilities, and thus, this is indeed another way for a person to find ready-to-run software. Future surveys should include this as a predefined answer choice.

To the extent that our results can be compared to those of Murphy-Hill et al. (2015), we find similarities and one difference. In the case of their participants' self-reports of tool discovery, peer recommendation was ranked lowest whereas in our survey, it was ranked second highest, which is a considerable difference. On the other hand, their "discussion thread" mode of discovery is comparable in meaning to our "Ask or search public mailing lists or discussion groups", and our results are nearly identical to theirs with around 21–22% of respondents reporting this as an approach they used; likewise, we also have very similar results for the use of Twitter and social media. We speculate that the difference in rating peer recommendations can be accounted for by the fact that their study focused on serendipitous discovery of software whereas here we asked about purposefully looking for software.

### 5.3. RQ2: What criteria do scientists and engineers use when choosing ready-to-run software?

As mentioned in Section 4.3, analysis of the software developer and nondeveloper responses to "In general, how important are the following characteristics when you are searching for ready-to-run software for a task?" did not indicate a statistically meaningful difference between the two groups; therefore, our analysis centers on the aggregate view of the responses (Figure 8).

The general trend of these results is, in many ways, what might be expected intuitively. For example, the highest-ranked criterion is the availability of specific features (96%) in the software—in fact, it was the only characteristic for which none of the respondents chose "Somewhat or occasionally important" or "Rarely or never important". This high ranking is unsurprising: after all, if one is searching for software for a task, paying attention to the software's feature set is paramount. Conversely, how software is implemented in terms of programming language (42%) and the particular software architecture (35%) were deemed relatively unimportant. This is also in line with expectations: if one is looking for ready-to-run tools, the details of the implementation often do not matter from a user's point of view, and instead, other operational constraints such as operating system support may be more important.

The results also show that support for specific data standards and file formats (93%) and software price (91%) are also major considerations. This may reflect the culture of scientific computing: software often is expected to be free, and specific areas of science often use specialized data formats. The apparent quality of the software (90%) also scored highly, as did operating system requirements (88%) and how easy the software is to learn (81%).

The middling rank of "other people's opinions" (64%) may seem surprising at first. In the responses to RQ1 (Figure 7), asking colleagues for opinions was chosen much more often (80% by developers, 92% by nondevelopers), so the results for the present question seem inconsistent. However, the explanation may be simple: RQ1 is about *approaches* to finding software, while RQ2 is more about *criteria* that people pay attention to when looking for software. In the latter context, people naturally must make judgements about what matters to them and their specific situation. While other people's opinions may be something that does indeed help people make a choice, it is a source of information that will probably be less important than other considerations such as availability of specific features in the software.

We can compare our aggregate results of Figure 8 to those of Lawrence et al. (2015) with respect to their question about important factors users consider when adopting new technology. There are notable differences. For example, in their survey, the highest-ranked factor was "documentation available", while in our survey, "quality of documentation" (the closest matching category) ranked sixth overall. Their second-ranked factor, "ability to adapt/customize" is close to our "how easy the software is to extend", which ranked seventeenth in our survey. While it is true that our survey included many more possible criteria, and in addition, some criteria in Lawrence et al.'s survey question were coarser in granularity, many items in both surveys are comparable, so these two differences alone are unlikely to explain the results. We hypothesize two possibilities. First, the context of their survey was scientific computing gateways, whereas our survey was not focused on this and considered people working with any kind of software environment. The contexts may influence the criteria people use. Second, it is possible that the rankings are influenced by the different answer formats: we asked participants to rank the importance of each criterion, while Lawrence et al. asked respondents pick their top three criteria.

Finally, some studies have examined the properties associated with successful open-source software projects (e.g., Crowston et al., 2003, 2006; Sen et al., 2012; Subramaniam et al., 2009; Tom Lee et al., 2009). How do those properties compare to the features that people in our survey indicated they used to discriminate between choices when looking for software? From among the most important traits found in the other studies (Crowston et al., 2003, 2006; Subramaniam et al., 2009; Tom Lee et al., 2009), we find code quality, documentation quality, price, and licensing terms are also in the top six of our Figure 8. Two other characteristics of successful open-source software projects—developers' reputations and other people's opinions—did not appear to be as important to our survey participants, based on the rankings of Figure 8. Nevertheless, the high ranking of four out of six properties suggests that future work could examine whether there is a causal relationship.

### 5.4. RQ3: What information would scientists and engineers like to find in a catalog of software?

Figure 9 summarizes the results for the question "Suppose that it were possible to create a public, searchable catalog or index of software, one that would record information about software of all kinds found anywhere. What kind of information would you find most useful to include for each entry in such a catalog or index?"

Analysis of the results shows that between software developers and nondevelopers in our sample of scientists and engineers, there is a statistically-significant difference in the preferences for only one characteristic: the name of the software. An additional characteristic, the purpose of the software, was ranked just as highly by nondevelopers (100%), but after applying correction for false positives, the difference in responses for developers and nondevelopers did not reach statistical significance for this characteristic. Still, it appears that respondents who are not involved in software development consider these two characteristics to be the most essential information to provide. This makes intuitive sense, since those constitute basic and essential information, and we are surprised that not all of the scientific programmers likewise indicated that information about the purpose of the software is essential. We have no hypothesis to explain this difference.

In terms of overal rankings of different characteristics, other details about the software, such as the types of user interfaces offered, a programmable API, and the programming language used to implement the software, were of middle importance to survey participants. It came as a surprise, however, that more formal indicators of software development rigor—such as test cases, well-commented code, and metrics evaluating code quality—ranked relatively low, even for developers. We expected developers to be more discerning about the quality of software they choose. A possible explanation is that developers may simply assume they will need to take a closer personal look at any software they choose and make their own judgement, so they don't regard it as important to include this information in a software index. This is another aspect of the results that would be worth investigating more deeply in future work.

Finally, ten individuals wrote text in the "Other" field of the question. Analysis of these responses revealed one answer was similar enough to the predefined categories that we included it in the counts shown in the graph, and one response was not interpretable. The remaining write-in values constituted additional categories of information that were not truly subsumed by any of the options we provided. The following are the distinct themes that were raised:

- Price (two mentions)
- Size of the user base (two mentions)
- Availability of documentation (two mentions)
- Size of the software
- Whether it is packaged for Debian
- URL of version control repository
- List of plug-ins available
- List of similar tools
- Stability of parent organization

*5.5. RQ4: How do software developers in science and engineering look for source code?*

The responses to this question revealed that the use of general search engines was the most popular approach (91%), followed by asking colleagues (53%), and in third place, a tie between consulting the literature and searching in repositories such as SourceForge, GitHub and BitBucket (45% each). This is consistent with findings in some other published studies (e.g., Lawrence et al., 2015), though the relative percentages are different.

The use of specialized software indexes such as ASCL.net ranked much lower (24%), as did searching code collections in one's organization (22%). Code search sites such as Open Hub ranked even lower (18%), and the use of social media systems such as Twitter, Facebook and LinkedIn ranked lower still (9%). Out of the four write-in "Other" answers, one was clearly in the same category as a predefined option, so we adjusted the counts accordingly; the other three were "O'Reilly books", "Look at the web page for that software!" and "What libraries are used by other software that I like?" These last three represent additional approaches not anticipated in our set of predefined answer choices.

These results show that close to half of respondents search project repositories such as GitHub when looking for source code, but unexpectedly, this approach is no more popular than looking in the scientific literature. This may reflect a population sample bias towards researchers in our study: *outside* of research environments, software developers may be less likely look in the research literature as often as they search in GitHub. On the other hand, we were surprised at the low ranking of searching topical software indexes.

How do these results compare to those for RQ1, which asked about finding ready-to-run software? Though similar, the two questions were not identical: three answer choices were different because the contexts lent themselves to some different actions, and in addition, the question from Figure 7 involved both developers and nondevelopers, whereas *this* question involved *only* those respondents involved in software development. Nevertheless, we can compare the common subset of answer options and the subset of respondents in Figure 7 who identified themselves as developers. We present the results in Figure 12.

The rankings in Figure 12 show that, overall, the top three approaches for finding *both* ready-to-run software and source code are identical: searching the Web, asking colleagues, and looking in the literature. When looking for source code, searching public repositories such as SourceForge and GitHub rises in popularity; while this is to be expected given the nature of the task and the fact that the respondents were software developers, the approach still only tied with searching the literature. At the other end of the rankings, the use of software indexes was still quite low overall (12% in the context of finding ready-to-run software, 22% in the context of source code). This result indicates that the developers of software catalogs continue to face challenges in producing systems that users find sufficiently compelling.

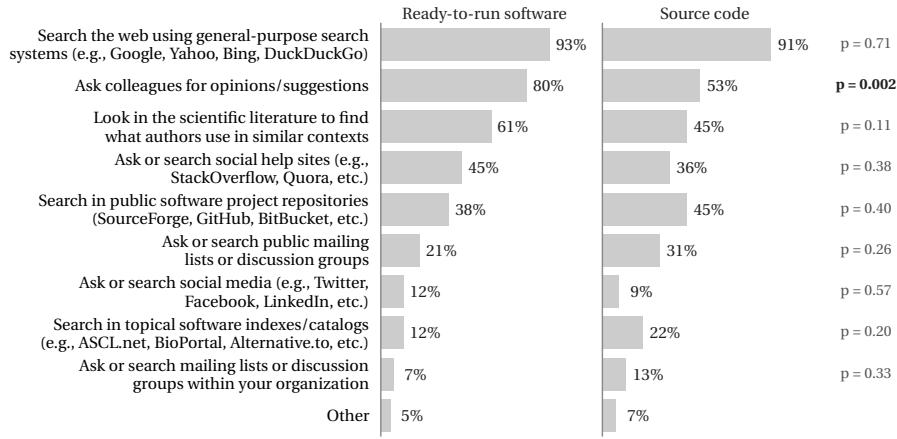

Figure 12: Comparison of the results from 7 and 10 for the overlapping answer categories. (Left) Subset of results from Figure 7 for the 56 respondents who indicated they were involved in software development. The results have been expressed as percentages of the total number of responses for that subgroup of people. (Right) Reproduction of the results of Figure 10. (Far right) P-values for the difference between results for each answer option. Bold typeface indicates results that are significant after applying the Benjamini-Hochberg procedure for controlling the false discovery rate to 10%.

Statistically, the only significant difference in responses is with respect to asking colleagues for opinions: it is far less common when searching for source code than when searching for ready-to-run software (53% versus 80%). We have no hypothesis to explain this difference; indeed, we find it unexpected and puzzling, and something that would be worth investigating in future research.

Some comparisons to other studies can be made. In their study of ways that security tools are discovered by software developers, (Xiao et al., 2014) asked 42 participants to indicate whether they had heard of tools via any of 14 predefined options (discussed in Section 2.4). Comparing their results to our Figure 12, we see there is some agreement in the results: their highest-ranked approach was coworker recommendation, which was ranked second in our results ("Ask colleagues for opinions/suggestions"), although the net percentage of our respondents who indicated this option was considerably higher in our case. The other comparable answer option in their study was "Online Forums and Discussion Boards", which is roughly comparable to our "Ask or search public mailing lists or discussion groups". Their respondents indicated this much less often than ours, however. Overall, we can conclude that both studies indicate peer recommendations are one of the most commonly-noted approaches to finding software.

*5.6. RQ5: What can prevent software developers in science and engineering from finding suitable source code?*

The results of our question about search failures (Figure 11) show that the largest hindrance is simply finding a match to one's needs, either because of difficulty finding suitable working software or because none of the options found satisfy requirements. Time limitations also often (46%) impact the ability to conduct proper searches for source code or to evaluate the results. This may be due to the large number of results that general-purpose search engines can return, which in turn may make it difficult to find suitable results easily.

The results also suggest that software licensing (12%) is rarely more than a minor hindrance, even though it was a more important criterion for ready-to-use software (Section 4.3). This suggests that intellectual property information is not sufficiently visible during searches. This is consistent with the format of results presented by Google and similar general-purpose search engines: they do not usually contain license information, unless it happens to be the in the first few words of the text fragment presented as part of a given search result.

Finally, one of the "Other" results written by respondents noted that some software packages lack web pages or other kinds of online presences. This is a notable observation. In effect, it means that the software is hidden from search engines, and may be hidden from search in social coding sites and social media as well. Unfortunately, we do not have data about the types of software in this category. Could it be that these "hidden" software packages are more likely to be older, noncommercial software? After all, commercial efforts are likely to seek to maximize publicity (in order to increase sales), while newer open-source efforts are likely to take advantage of online systems such as GitHub. This is a question that could be probed in future surveys.

Overall, our results are very similar to those of Samadi et al. (2004). For our question about what factors hindered scientists and engineers from finding software (Section 4.6), the most popular reason was finding a match to one's needs. On the single point of intellectual property issues (e.g., licensing), our results also align with those of (Frakes and Fox, 1995). In that study, they asked the question "I'm inhibited by legal problems", which is subsumed by one of our answer options for RQ5. They found that legal problems did not inhibit code reuse, which corresponds to our finding that people rarely listed licensing issues as a problem.

## 6. Conclusions

Before the advent of the World Wide Web, even before the current Internet, it was arguably easier to find existing software for personal computers—there was less of it, and there were simply fewer places to look. Community bulletin boards and archive sites using FTP made software available for copying by anonymous users over telephone networks; later, the Usenet culture (Emerson, 1983) of the 1980's encouraged widespread sharing and even devoted a newsgroup (*comp.sources*) to the exchange of software source code. Communities created manually-curated lists of software (e.g., Boisvert et al., 1985; Brand, 1984) and some journals regularly published published surveys of topical software (e.g., Martinez, 1988). The breadth of software we have today did not exist then, but one could feel reasonably sure to have found and examined the available options in a finite amount of time. Fast-forward to today, and the staggering wealth of software resources available to users is both a blessing and a curse: one can simultaneously feel that for any given task, "surely *someone* has already written software to do this," and yet an attempt to find suitable software can seem like falling into a rabbit hole.

So what *do* users do today when they want to find software? This survey was an attempt to gain insight into the approaches used by people working in science and engineering, including criteria that they apply to select between alternative software choices. Our participants were experienced researchers worked primarily in the physical, computing, mathematical and biological sciences; the majority were involved in software development and had a mean of 20 years of experience; most worked in small groups; and all had some degree of choice in the software they used. The majority spent over 50% of their day using software; this is somewhat higher than some other studies have reported (e.g., Hannay et al., 2009, found scientists spent 40% of their time using scientific software).

The survey results help identify a number of current community practices in searching for both ready-to-use software and source code:

1. When searching for ready-to-run software (RQ1), the top three approaches overall were are: (i) search the Web with general-purpose search engines, (ii) ask colleagues, (iii) look in the scientific literature. After these top three, the next most commonly stated approaches differed between those respondents who self-identified as being involved in software development and those did not: more developers in our sample indicated asking on social help sites such as Stack Overflow and searching in public software repositories such as GitHub (in that order), while nondevelopers indicated following their organization's guidelines and a tie between asking on public mailing lists and asking on social media. We found statistically significant differences between the subgroups' uses of social help sites, software project repositories, software catalogs, and organization-specific mailing lists or forums.

2. Our RQ2 revealed that the top five criteria given above-average weight when searching for ready-to-run software are: (i) availability of specific features, (ii) support for specific data standards and file formats, (iii) price, (iv) apparent quality of the software, and (v) operating system requirements. On the other hand, the least important criteria were (a) size of software, (b) software architecture, and (c) programming languages used in implementation.

3. Regarding information that workers in scientific and engineering fields would like to see in a catalog of ready-to-run software, a total of 15 features were indicated as having above-average value by at least 50% of the respondents; of these characteristics, the operating system supported, purpose of software, name of software, domain/field of application, and licensing terms were the five most-often requested features. Software developers different from nondevelopers in our sample of scientists and engineers in that they rated the name and purpose of the software as the most important information to provide. Another slight difference involved information about the availability of support or help for a given software product, but on the whole, both subgroups displayed similar preferences.

4. The top five approaches used by software developers in science and engineering to search for source code are almost identical to those they use to find ready-to-run software. They are: (i) search the Web with general-purpose search engines, (ii) ask colleagues, (iii) look in the scientific literature, (iv) search in public software project repository sites such as GitHub, and (v) look in social help sites such as Stack Overflow.

5. The top three reasons given by the developers in our sample for why they were sometimes *unable* to find source code are: (i) unable to locate suitable software, (ii) requirements are too unique, and (iii) insufficient time to search or evaluate options. Conversely, concerns about intellectual property issues ranked low.

The results above have implications for the development of better resources for locating software. In common with other surveys, we found that more people indicate they use general Web search engines than any other approach for finding both ready-to-run software and source code. This implies that for any specialized resource such as a software catalog to gain popularity, it must be indexed by Google and other search engines so that users can find its content via general Web searches. Our results for RQ2 (Figure 8) also point out information that people consider important when looking for software; this can be used to inform the development of more effective software search systems. For example, if one were creating a software search engine, providing direct access to information about data formats supported by different software tools could help scientists and engineers to find and select tools more quickly. Finally, software cataloging efforts would benefit by focusing on presenting the most desirable information revealed by RQ3 in our survey (Figure 9).

Though our survey considered only general resources, there also exist a number of source code finding systems today integrated into specialized software development tools (e.g. Hoffmann et al., 2007; Linstead et al., 2009, 2008; Martie et al., 2015; Ossher et al., 2009; Ye and Fischer, 2002; Zagalsky et al., 2012). Software developers can take advantage of these systems to find software during development activities. Though our survey did not specifically examine the use of these tools, we would expect that the attributes rated most important in Figure 8 would also be relevant in the context of using such integrated code-finding facilities. However, this hypothesis is untested, and constitutes a question that future studies could explore.

### 6.2. Lessons for future surveys

Analyzing the survey results has led us to recognize aspects of the survey that could have been improved. First, in the demographic profile questions (Section 4.1), it would have been useful to gather more specific data. For example, the work fields question could have offered finer-grained options, and additional questions could have asked participants about their institutional affiliation (e.g., educational, government, industry) as well as their work roles (e.g., student, staff, faculty). Of course, the benefits of additional questions must be weighed against respondents' patience for filling out long surveys.

Second, the questions asking about software search could have had an ex-plicit answer choice about the use of scientific gateways. The survey questions generally did not mention gateways or portals explicitly; the closest was the question discussed in Figure 9, which included workflow environments as an answer choice. Based on the responses reported in Figure 9, one quarter of the respondents consider support for workflow environments a criterion in selecting software. Since we did not ask about it explicitly, it is unclear whether any of the participants had the use of gateways in mind and framed their responses accordingly. It is also not clear what effect this would have had on their responses. Gateways concentrate software resources in one location and typically provide an index or other means of finding software provided by the gateway, and it is conceivable that this may change the nature of how users think of finding software or the criteria they use to discriminate between available alternatives. It is therefore possible that this is a confounding factor in our results. Future surveys should address this aspect explicitly.

Third, future work must strive to increase the response rate. While we believe the present survey's results are accurate for the sample of people who finished the survey, we must also acknowledge that a response rate of 3% is disappointing. It is widely asserted that Web-based surveys often encounter low rates (e.g., Couper, 2000; Couper and Miller, 2008; Kitchenham and Pfleeger, 2008); in our experience, many studies even fail to disclose the response rate, or claim a rate without reporting the number of potential recipients, leaving in question the accuracy of the rate. However, of the published surveys that disclose both the number of potential recipients and the number of completed responses received (e.g., Bauer et al., 2014; Kalliamvakou et al., 2014; Lawrence et al., 2015; Sojer, 2010; Wu et al., 2007), the values often have been higher. For example, Sojer (2010) reported 9.7% and Lawrence et al. (2015) obtained 17%, albeit with a highly motivated population. One possible cause for our lower response rate may be the venues where we advertised the survey. Our primary venues for soliciting participation were certain mailing lists and Facebook groups. With respect to the mailing lists, some recipients may not have received the survey messages because automatic spam filters may have blocked the mes-sages from their electronic mail inboxes. This would mean that fewer people saw the invitations than the number of people subscribed to the mailing lists, artificially reducing the apparent response rate. With respect to Facebook, some users may be have signed up long ago but they may rarely or never check the group we targeted. The latter is especially plausible when we consider two other results of our survey: as shown in Figure 4, respondents had a mean of 20 years of experience, and in Figure 12, social media of Twitter/Facebook/LinkedIn variety were little-used by participants for finding software. If that reflects the overall population we reached and their broader pattern of social media use, then they may simply be of a generation that spends less time on Facebook than a younger generation of researchers. Again, this would cause our estimated number of recipients to be higher than the actual number of people who saw the announcements in that venue. Finally, it is possible that our announcements and/or the front page of the survey were simply not sufficiently motivational.

## 7. Acknowledgments

We thank Alice Allen, Daniel S. Katz, Sarah M. Keating, Matthias König, Allyson Lister, Cristina V. Lopes, Chris Mattmann, Rajiv Ramnath, Renee M. Rottner, Lucian P. Smith, and Linda J. Taddeo for many comments and feedback on previous versions of this manuscript and the survey. We also thank two anonymous reviewers for insightful comments and challenging questions that ultimately improved the manuscript. This work was supported by the USA National Science Foundation (award #1533792).